\documentclass[review]{elsarticle}

\usepackage{lineno}
\modulolinenumbers[5]

\journal{Journal of \LaTeX\ Templates}

\usepackage{graphicx}
\usepackage{natbib}
\usepackage{multirow}
\usepackage{color}
\usepackage{wrapfig,framed,caption,booktabs}
\usepackage{amsmath}
\usepackage{ulem}
\usepackage{soul}

\usepackage{hyperref}
\hypersetup{colorlinks=true,linkcolor=red,citecolor=blue}

\pdfoutput=1

\newcommand{\arcmin}{\hbox{$^\prime$}}
\newcommand{\arcsec}{\hbox{$^{\prime\prime}$}}
\newcommand{\utw}{\smash{\rlap{\lower5pt\hbox{$\sim$}}}}
\newcommand{\udtw}{\smash{\rlap{\lower6pt\hbox{$\approx$}}}}

\newcommand{\farcs}{\hbox{$.\!\!^{\prime\prime}$}}

\begin{document}

\begin{frontmatter}

\title{Design of an adaptable Stokes polarimeter for exploring chromospheric magnetism}

%%\titlerunning{Adaptable Stokes polarimeter for exploring chromospheric magnetism} 

\author[mymainaddress,mysecondaryaddress]{Rohan E. Louis\corref{mycorrespondingauthor}}
\cortext[mycorrespondingauthor]{Corresponding author}
\ead{rlouis@aip.de}

\author[mythirdaddress]{A. Raja Bayanna}

\author[mysecondaryaddress,myfourthaddress]{H\'ector Socas Navarro}

\address[mymainaddress]{Leibniz-Institut f\"ur Astrophysik Potsdam (AIP), 
                        An der Sternwarte 16, 14482 Potsdam, Germany}
\address[mysecondaryaddress]{Instituto de Astrof\'isica de Canarias, 
                             E-38200 La Laguna, Tenerife, Spain}
\address[mythirdaddress]{Udaipur Solar Observatory, Physical Research Laboratory, 
                         Dewali, Badi Road, Udaipur, Rajasthan 313004, India}
\address[myfourthaddress]{Departamento de Astrof\'isica, Universidad de La Laguna, 
                           La Laguna, Tenerife E-38205, Spain}

%%\authorrunning{Louis \textit{et al.}} 

\begin{abstract}
The chromosphere is a highly complex and dynamic layer of the Sun, that serves as a conduit for mass 
and energy supply between two, very distinct regions of the solar atmosphere, namely, the photosphere and 
corona. Inferring magnetic fields in the chromosphere, has thus become an important topic, that
can be addressed with large-aperture solar telescopes to carry out highly sensitive polarimetric 
measurements. In this article, we present a design of a polarimeter 
for investigating the chromospheric magnetic field. The instrument consists of a number of lenses, 
two ferro-electric liquid crystals, a Wollaston prism, and a CCD camera. The optical design is similar 
to that of a commercial zoom lens which allows a variable f\# while maintaining focus and aberrations
well within the Airy disc. The optical design of the Adaptable ChRomOspheric POLarimeter (ACROPOL) 
makes use of off-the-shelf components and is described for the 70~cm Vacuum Tower Telescope and the 
1.5~m GREGOR telescope at Observatorio del Teide, Tenerife, Spain. Our design shows that the optical 
train can be separated into two units where the first unit, consisting of a single lens, has to be 
changed while going from the VTT to the GREGOR configuration. {\color{black}We also discuss the 
tolerances within which, diffraction limited performance can be achieved with our design.}
\end{abstract}


\end{frontmatter}

%%\linenumbers

\section{Introduction}
\label{intro}
The chromosphere couples the photosphere to the corona. It represents an important region 
in the solar atmosphere where nearly all of the mechanical energy that drives solar activity 
is converted into heat and radiation, with only a small fraction leaking through to heat the 
overlying corona and power the solar wind \citep{1977ARA&A..15..363W}. Several 
processes, such as wave propagation and dissipation \citep{2001ApJ...561..420M}, 
electrical currents \citep{2008ApJ...687.1388J}, and magnetic reconnection 
occur in the chromosphere, which regulate the mass and energy supply to the corona. 
However, the relative contribution of the individual processes, their dependence on local conditions, 
and the nature of non-thermal to thermal energy conversion, remains an open question. Thus, 
understanding the chromosphere and its magnetism is a significant necessity for explaining 
the corona and heliosphere \citep{2014SoPh..289.2733D}. 

As the pressure scale height in the solar atmosphere is about 150~km 
\citep{1988atsu.book.....D}, 
the chromospheric density is about four orders of magnitude lower than that in the photosphere. 
Subsequently, the Alfv\'en and sound speeds in the chromosphere necessitate measurements at high 
temporal and spatial resolution so as to resolve gas, plasma, and wave motions 
\citep{2012SPIE.8443E..4FK}. Diagnosis of the chromospheric magnetic field is 
mainly done through spectro-polarimetry in spectral lines formed under non-LTE 
conditions \citep{2000ApJ...530..977S}. More importantly, the chromospheric 
signature in these spectral lines is only evident in a narrow range around the line core, where 
the photon flux is only 10-20\% of the continuum intensity. A polarimetric sensitivity of the 
order of 10$^{-4}$ is required to detect weak chromospheric fields 
\citep{2010AN....331..581S}. 
The extraction of the chromospheric 
magnetic field at high spatial and temporal resolution is quintessential to shed light into 
this complex region of the Sun and requires large-aperture solar telescopes with 
photon-efficient instruments. While the 4-m Daniel K. Inouye Solar Telescope 
\citep[DKIST, previously ATST;][]{2008AdSpR..42...78R}, and the 4-m European Solar Telescope 
\citep[EST;][]{2013MmSAI..84..379C} are important steps in this direction, the 
former is presently under construction and it will be several years until the latter is realised. As 
Europe's largest operational solar telescope, the 1.5-m 
GREGOR \citep{2012AN....333..796S}, at Observatorio del Teide, is pivotal for exploring 
chromospheric magnetism and on the verge of creating several important scientific milestones 
which will serve as a foundation for future large-aperture solar telescopes. 

A design for a dual-beam polarimeter is presented in this article which will probe 
the chromospheric magnetic field using the Ca~{\sc ii} infrared line at 854.2~nm. 
The aim of this work is to determine the feasibility of designing a single optical 
configuration that can modify the f\# at the detector plane, analogous to commercial zoom lenses. 
In this article we discuss the design specifically for different input f-numbers corresponding
to the 70~cm Vacuum Tower Telescope (VTT) and the 1.5~m GREGOR.

The organisation of this article is as follows: diagnostics for chromospheric spectro-polarimetry and 
some key scientific questions which could be addressed, are presented in Sects.~\ref{lines} and 
\ref{science}, respectively. The instrument and its performance are described in Sect.~\ref{instr}, 
and the concluding remarks are highlighted in Sect.~\ref{conclu}.

\section{Diagnostics for chromospheric spectro-polarimetry}
\label{lines}
There are a number of spectral lines that can be used for chromospheric observations. The Hydrogen 
H$\alpha$ Balmer line at 656~nm is historically significant and has been traditionally used for flare, 
spicule, and prominence/filament studies. Despite this, its interpretation is enormously challenging 
as the population of the excited $n=3$ level is strongly dependent on the temperature and radiative 
conditions in the solar atmosphere, with quantitative estimates confined to simplistic assumptions 
such as the cloud model \citep{1964PhDT........83B}. Other prominent spectral lines are the Ca {\sc ii} 
H\&K lines at 390 nm, the NaD line pair at 589~nm, the Mg~{\sc i}~B2 at 517~nm, the He~{\sc i}~D3 at 
588~nm, the Ca~{\sc ii} infrared triplet at 850~nm, and the He~{\sc i} infrared triplet at 1083~nm. The 
Ca~{\sc ii} and He~{\sc i} infrared lines have gained considerable importance over the last decade due 
to sensitive polarimeters (references provided below) and the availability of 
standard inversion codes \citep{2008ApJ...683..542A,2009ASPC..415..327L,2015A&A...577A...7S} that 
provide physical parameters in the relevant part of the solar atmosphere. 

The He~{\sc i} 1083~nm line forms through ionisation by coronal ultra-violet (UV) radiation, thus 
forming in a narrow region in the upper layers of the chromosphere 
\citep{1956ApJ...124...13M,1977SoPh...54..343H,1992SoPh..138..107V,1993ApJ...406..319F}. 
While the line is free from photospheric contamination, it is extremely weak in the quiet Sun. 
The Ca~{\sc ii} infrared triplet on the other hand, has a low excitation potential of 1.69~eV 
and extends from the upper photosphere well into the chromosphere. 
For instance, at $\pm60$~pm from the line core, one can 
probe the photosphere at a height of about 200~km which exhibits reversed granulation. The ``knees'' 
of the intensity profile at around $\pm30$~pm sample the temperature minimum region, some 500~km above 
the continuum forming layer. The line core, representing heights of approximately 1300~km, is purely 
chromospheric \citep{2009ApJ...694L.128L}. Thus, the Ca~{\sc ii} infrared line 
has the advantage, 
that it can be combined with photospheric observations to extract the stratification of physical 
parameters over a large height 
range in the atmosphere. The sensitivity of the Ca~{\sc ii} infrared line to longitudinal and transverse
magnetic fields has been recently studied by \cite{2016MNRAS.459.3363Q} and the authors conclude that 
the line is mostly sensitive to the layers between $\log{\tau}=0$ and $\log{\tau}=-5.5$, under the prescribed
conditions in their model atmospheres.

Several instruments at various ground-based telescopes observe the He~{\sc i} infrared triplet. They 
are the Spectro-Polarimeter for Infrared and Optical Regions \citep[SPINOR;][]{2006SoPh..235...55S}, the 
Facility Infrared Spectropolarimeter \cite[FIRS;][]{2010MmSAI..81..763J}, and the SOLIS Vector 
SpectroMagnetograph \citep[VSM;][]{2006ASPC..358...92H} at the National Solar Observatory (NSO), and the  
The Near-InfraRed Imaging Spectropolarimeter \citep[NIRIS;][]{2012ASPC..463..291C} at the 
Big Bear Solar Observatory. Observations in the Ca~{\sc ii} infrared line are done with the 
Interferometric BIdimensional Spectrometer \citep[IBIS][]{2006SoPh..236..415C} and SPINOR at NSO, 
as well as with the CRisp Imaging SpectroPolarimeter \citep[CRISP;][]{2008ApJ...689L..69S} at the 
Swedish Solar Telescope. 

The 1.5-m GREGOR Infrared Spectrograph \citep[GRIS;][]{2012AN....333..872C} is routinely providing 
observations of the He~{\sc i} infrared triplet, in continuation to the Tenerife Infrared 
Polarimeter \citep[TIP-I/II;][]{2007ASPC..368..611C}, which was operational at the adjacent 
70~cm VTT for more than a decade. Our optical design for the polarimeter, 
allows it to be mounted at the VTT or GREGOR, which will provide a comprehensive coverage of the solar
chromosphere and significantly reinforce the instrumentation capability of the two telescopes. 

\section{Scientific outlook}
\label{science}
In this section, the usefulness of our instrument is briefly described for a few research areas in solar physics.

\subsection{The 3D structure of sunspots} 
Understanding the thermal, magnetic, and kinematic structure of sunspots has been one of the frontier 
topics in solar physics. While the magnetic field can be inferred with high spatial resolution in the 
photosphere owing to large telescopes and sensitive polarimeters, the vector magnetic field is only 
known at a single height or as a function of optical depth. \citet{2010ApJ...720.1417P} carried out 
a geometrical transformation for a small penumbral region, allowing physical quantities such as the 
electrical current density, helicity, Wilson depression, plasma $\beta$, etc. to be derived. However, 
to do the same for a full sunspot, one requires inverting spectral lines spanning a large height range 
in the solar atmosphere, so as to satisfy the condition of force balance between structures such as 
the umbra and penumbra, that are relatively depressed by about 300~km \citep{2004A&A...422..693M}. 
The above underscores the importance of the chromospheric magnetic field to extract basic physical 
quantities associated with sunspots.

\subsection{Spectro-polarimetry of PILs in complex active regions}
Solar eruptions can be triggered through the emergence of flux which occur when newly emerged magnetic 
fields appear in a region of pre-existing flux, often in, or in close proximity to, a filament channel 
\citep{1972SoPh...25..141R}. This can lead to the formation of a magnetic flux rope through bodily 
emergence \citep{1996SoPh..167..217L}, by reconnection within an emerged magnetic arcade 
\citep{2004ApJ...610..588M}, or by reconnection with the pre-existing flux \citep{2012ApJ...760...31K}. 
The source regions comprise highly sheared magnetic fields in the corona which overlie a photospheric 
polarity inversion line \citep[PIL;][]{1968SoPh....5..187M,1990ApJS...73..159H}. These locations 
inevitably show a filament channel in the chromosphere and often contain a filament or prominence 
in the corona above \citep{1998SoPh..182..107M}. It is widely (but not universally) accepted that 
a filament represents a weakly twisted magnetic flux rope holding the cool material 
\citep{2010SSRv..151..333M}. The formation and instability of a flux rope is a key element of 
storage-and-release eruption models 
\citep{1978SoPh...59..115V, 1995ApJ...446..377F,2006ApJ...641..577M, 2006PhRvL..96y5002K}.

Multi-wavelength observations offer the possibility to study the formation of flux ropes at PILs 
through the process of flux emergence, especially in the vicinity of active regions 
\citep[ARs;][]{2014A&A...562A.110L,2015SoPh..290.3641L}. The formation of PILs in $\delta$-spots 
\citep{2014A&A...562L...6B}, in particular, is highly interesting and important because of 
the presence of strong magnetic fields that can store large amounts of free magnetic energy, which 
can be impulsively released at the time of flares. Combining the photospheric and chromospheric 
magnetic field allows for accurate estimates of the free magnetic energy which can help constrain 
existing flare models and in determining the processes which destabilise the magnetic field at PILs.

\subsection{Establishment of chirality in AR filaments} 
Chromospheric H$\alpha$ observations often show filaments having a magnetic pattern of handedness or 
``chirality''. When the axial field is directed rightward, when viewed by an observer at the 
positive-polarity side, the filament is said to be `dextral' \citep{1997SoPh..175...27Z}. On the 
other hand, if the axial field is directed leftward from the same perspective, the filament is called 
`sinistral'. \citet{2000ApJ...540L.115C} showed that filament chirality bears a close correspondence 
to the sign of magnetic helicity in ARs, wherein sinistral and dextral filaments had positive and 
negative magnetic helicity, respectively. Classifying filaments as sinistral or dextral requires 
identifying the orientation of bright and dark filament threads from EUV images. A combination of 
localised heating and energy transport along the field lines could render filament threads to 
appear in emission, which can only be confirmed using chromospheric spectro-polarimetry. Thus, it 
is necessary to determine the boundary conditions and small-scale instabilities that influence 
filament chirality which would be critical for understanding the relation of erupting filaments 
to CMEs, as well as to the magnetic structure and formation of AR filaments themselves.

\begin{figure*}[!h]
\centerline{
\includegraphics[angle=0,width=\textwidth]{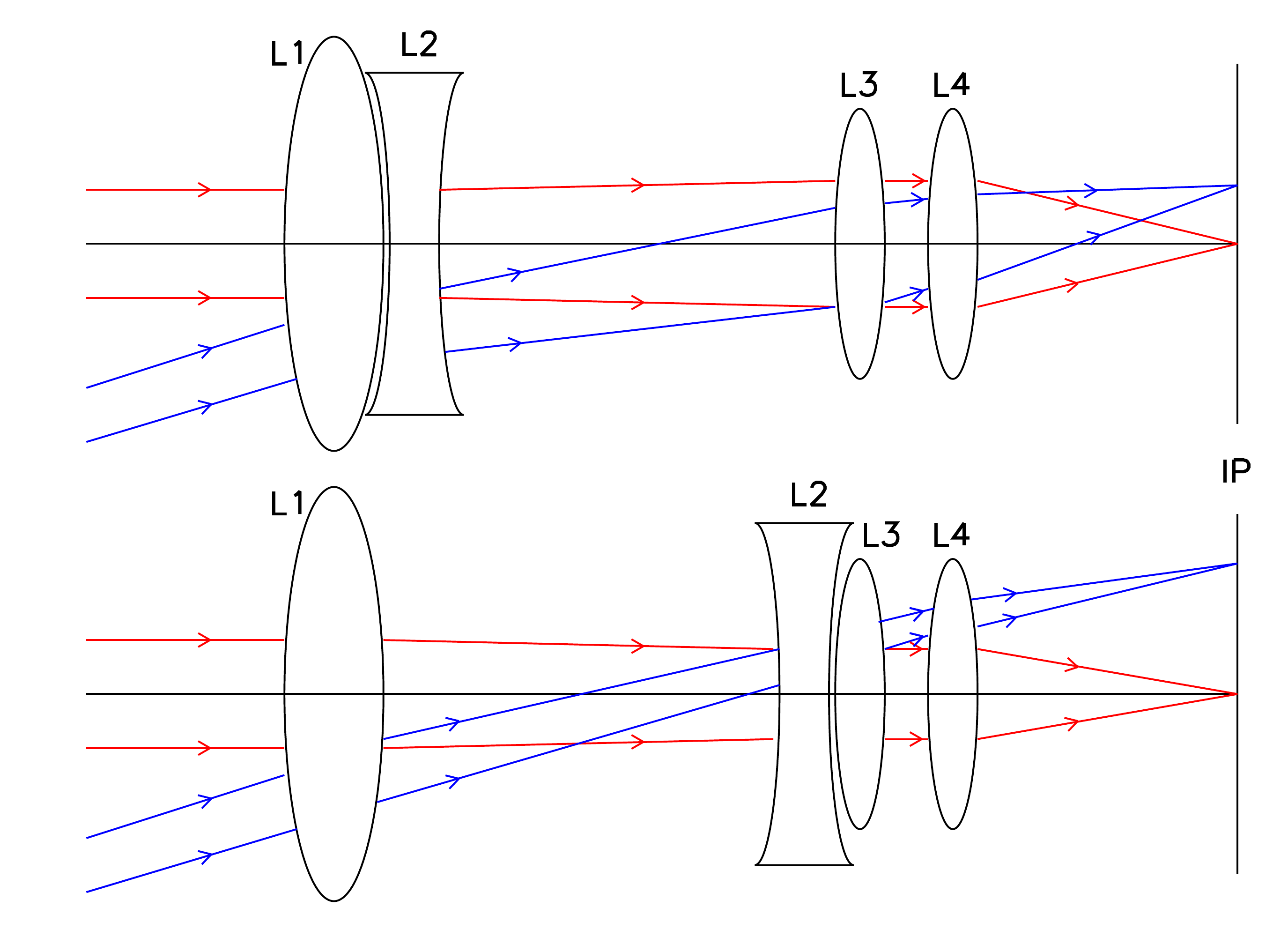}
}
\caption{Illustration of zoom lens principle. The zoom lens consists of several lenses, some of
which can be moved back and forth to change the magnification. In this simple example the 
magnification is changed by moving the second lens L2 while maintaining focus at the image plane 
(IP). The red and blue lines correspond to different field points.}
\label{fig01}
\vspace{-50pt}
\centerline{
\includegraphics[angle=180,width=\textwidth]{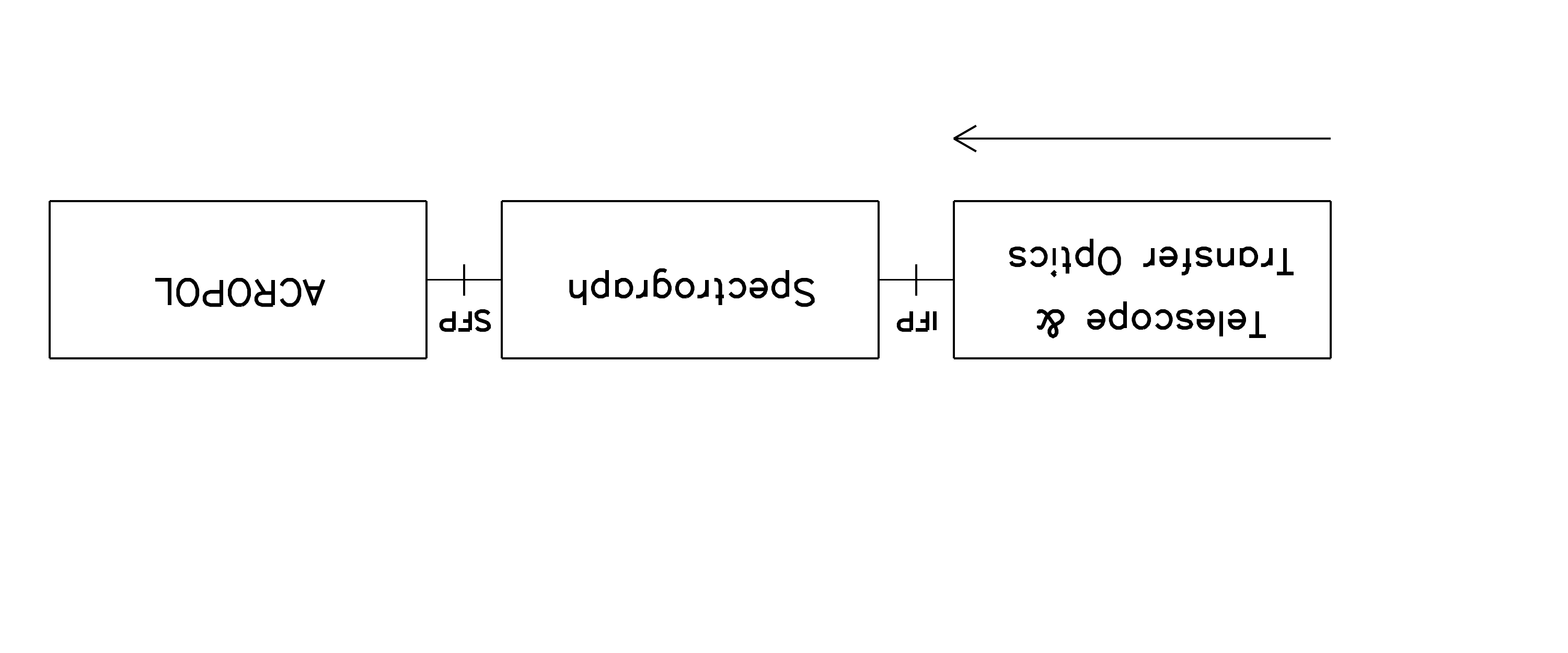}
}
\vspace{-35pt}
\caption{Block diagram illustrating the placement of ACROPOL in a general optical setup. The arrow
indicates the direction of light. IFP and SFP refer to the instrument focal plane and spectrograph 
focal plane, respectively.}
\label{fig01a}
\end{figure*}

\section{Adaptable ChRomOspheric POLarimeter (ACROPOL)}
\label{instr}
The plate scale for post-focus instruments is determined by the telescope diameter and the pixel
size of the detector. In order to convert the f\# at the instrument plane to that at the detector,
a pair of relay lenses are often used, where one serves as the collimator and the other as the imager.
As is evident, this configuration is unique to the telescope diameter and f\#, assuming the detector
size remains unchanged. The concept of an Adaptable ChRomOspheric POLarimeter (ACROPOL) was 
introduced to explore the possibility of a design, wherein the above limitation could be 
overcome, with a single post-focus instrument. In other words, the optics in the instrument operate
much the same way as commercial zoom lenses which allow a variable f\# and plate scale. Fig.~\ref{fig01}
shows the working of a simple zoom lens, in which the movement of one of the lenses 
(lens L2 in the figure) allows 
the f\# to change while keeping the focal plane unaltered. Such a system is referred to as a 
parfocal system. Along with a number of lenses, ACROPOL comprises
polarizing optics -- two Ferroelectric Liquid Crystals (FLCs) as polarization modulators and a 
Wollaston prism (WP) to produce the orthogonal polarized components, and a CCD camera as the 
detector. 

A number of requirements had to be met by the instrument. These included,  
diffraction limited performance, using off-the-shelf components, adaptability to an existing set-up, 
compactness, portability, and cost effectiveness. While ACROPOL is intended to operate at any 
existing solar facility, the design study presented here, was carried out for the GREGOR and VTT. 
Figure~\ref{fig01a} shows the block diagram of the setup where ACROPOL will be 
deployed. This layout is common for the both the VTT and the GREGOR. The instrument will be mounted 
at the spectrograph focal plane (SFP in Fig.~\ref{fig01a}), thus complementing the GREGOR 
Fabry-P\'erot Interferometer \citep[GFPI;][]{2012AN....333..880P} and the Triple Etalon SOlar 
Spectrometer \citep[TESOS;][]{2002SoPh..211...17T}, but nothing 
prevents it from being used at the focal plane of a 2-dimensional imaging spectrometer. 
The choice of using ACROPOL with a spectrograph is to keep the interfacing 
with the image scanning mechanism as simple and minimal as possible. We now proceed to describe the 
functionality of the individual components of ACROPOL.

\subsection{Optics}
\begin{figure*}[!ht]
\vspace{-20pt}
\centerline{
\hspace{-15pt}
\includegraphics[angle=180,width=1.15\textwidth]{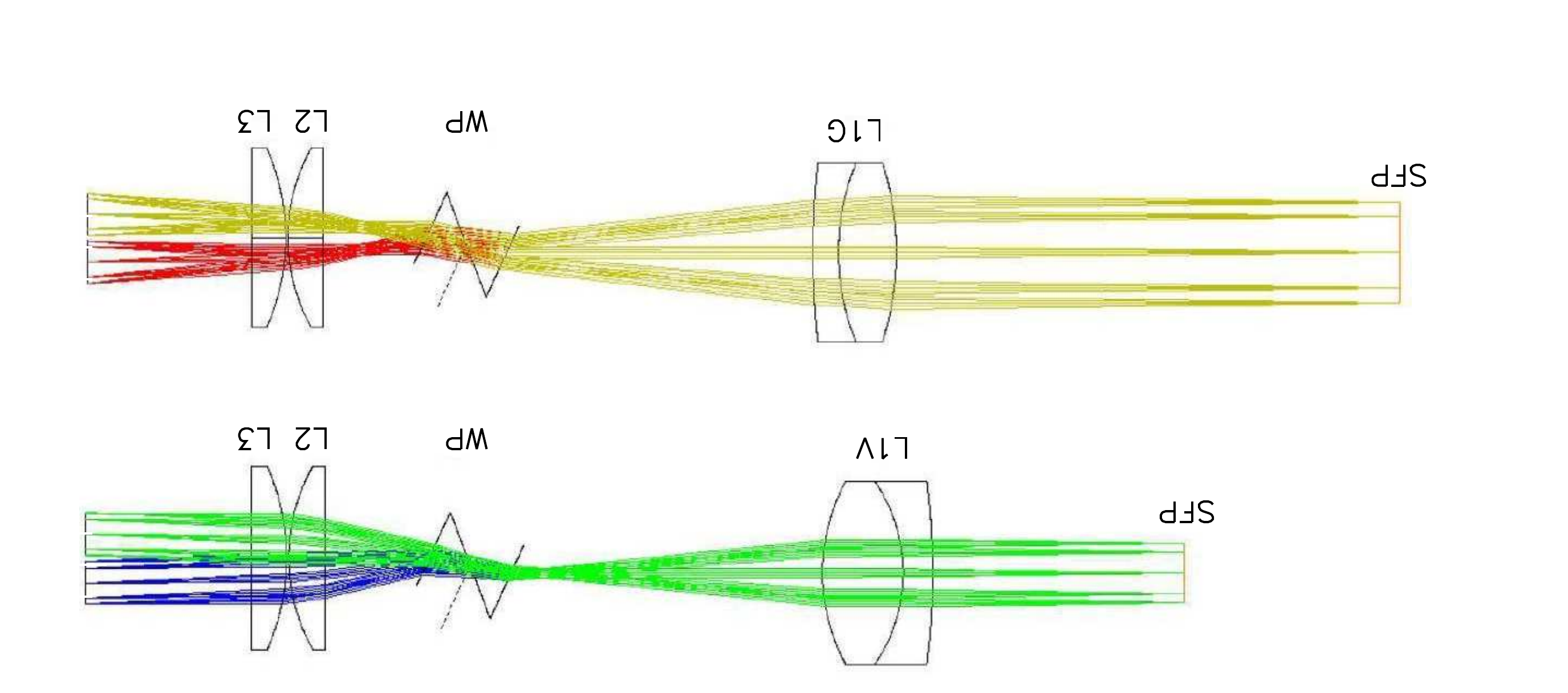}
}
\vspace{-20pt}
\caption{Ray diagram of ACROPOL for the VTT (above) and GREGOR (below) configurations depicting,
dual-beam polarimetry. The incident beam enters the optical train at the spectrograph focal plane 
(SFP) on the left. The ray propagation is along the $z$-axis and the $x$-axis is perpendicular 
to the plane of the paper. In the figure the wedge angle of the WP is rotated by 24.5$^\circ$ 
to produce the required separation of the ordinary and extra-ordinary rays and to avoid change 
in f\# of the O and E rays. The rotation of the wedge angle w.r.t to the $y$-axis is only for 
illustration, whereas during alignment the WP will be rotated w.r.t to the $x$-axis 
for mounting purposes.}
\label{fig02}
\end{figure*}

The optical design of ACROPOL is intended for a range of plate scales without compromising 
the image quality, and the parfocal design should ensure that the focal plane of the detector 
remains unchanged. Table~1 summarises, the telescope parameters and the design 
requirements. The inputs for the design are the wavelength of interest, field-of-view (FOV), 
magnification, detector size, and minimum separation between the two beams. The optical design 
was carried out using ZEMAX$^{\textrm{\tiny\textcopyright}}$ for 854$\pm$1~nm and the ray 
diagrams are shown in Fig.~\ref{fig02}. The design includes two configurations, one changing the 
f\#33 beam to f\#23 for the VTT, and the other converting an f\#40 beam to an f\#17 beam for GREGOR.

\begin{table}[!h]
\begin{center}
{\renewcommand{\arraystretch}{1.1}
\normalsize
\scalebox{0.8}{
\begin{tabular}{l|c|c}
{\textbf{Telescope and Image Dimensions}}           & VTT          &     GREGOR \\
\hline
Telescope diameter (mm)                             & 700          & 1440      \\
f\# at spectrograph slit                            & 66           & 40      \\            
Mag. between coll. and camera mirror                & 0.5          & 1      \\
f\# at instrument focal plane                       & 33           & 40      \\
FOV (~\arcsec)                                      & 72           & 50      \\     
Size of above FOV at instrument focal plane (mm)    & 8.07         & 14.04   \\                         
f\# at CCD focal plane                              & 23           & 17 \\
Diffraction limit at 8542~\AA~(~\arcsec)            & 0.31         & 0.15   \\
Image size on CCD (mm)                              & 12.5         & 12.7   \\
Spatial sampling at CCD (~\arcsec$/$pixel)          & 0.15         & 0.102  \\
Separation between two beams (mm)                   & 0.85         & 0.6    \\
\hline
&&\\
{\textbf{Dimensions of Optical Components}}         &              &   \\
L1 focal length (mm)                                & 40           &  63 \\
Edmund Optics                                       & 45801        &  49797      \\
L2 focal length (mm)                                & 50           &  50 \\
Edmund Optics                                       & 32478        &  32478  \\
L3 focal length (mm)                                & 60           &  60  \\
Edmund Optics                                       & 45128        &  45128  \\
Tilt of WP ($^\circ$)                               & 24.5         &  24.5 \\             
WP (mm)                                             & 11$\times$11 &  11$\times$11 \\
Thorlabs                                            & WPA10        &  WPA10  \\
Distance between instrument focal plane and L1 (mm) & 34.5         &  64.7 \\
Combined focal length of L2 \& L3                   & 27.5         &  27.5 \\ 
Distance between L1 and WP                          & 43           &  43 \\
Distance between WP and L2/L3                       & 11           &  11 \\
Distance between L2/L3 and sensor                   & 22.7         &  22.7 \\
\hline
\end{tabular}}
}
\caption{Image dimensions and f\# conversion using ACROPOL at the VTT and GREGOR. The stock 
numbers for the optical components have also been included.}
\label{tab01}
\end{center}
\end{table}

The telecentric beam from the spectrograph (SFP in Fig.~\ref{fig02}) is re-imaged onto the detector using 
a number of lenses. A Wollaston prism (WP) is placed in the optical path to obtain both states 
of polarization simultaneously. Initially, we tried to use four or five lenses to vary the f\# 
for different configurations, by systematically adjusting the distance between 
the individual lenses. 
However, due to the small size of commercial off-the-shelf Wollaston prisms, and the separation required 
between the extraordinary (E) and ordinary (O) beams, diffraction limited performance 
could not be achieved. In the present design, diffraction limited performance could be 
achieved with only three lenses. However, one of the lenses has to be changed between the VTT and 
GREGOR configurations, while the position of all the remaining components remains unaltered.

\begin{figure*}[!h]
\vspace{-90pt}
\centerline{
\includegraphics[angle=90,width=1.2\textwidth]{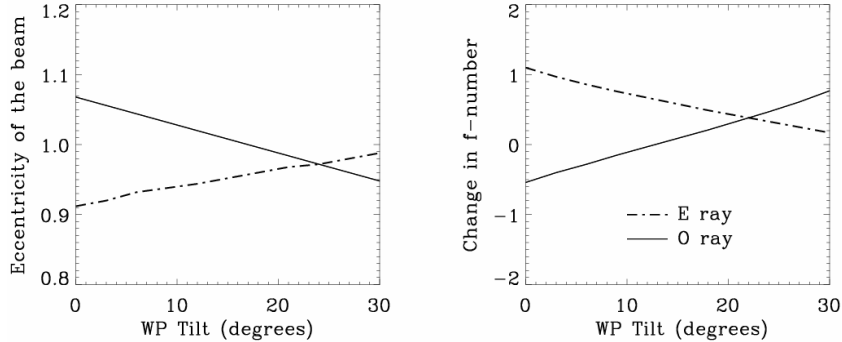}
}
\vspace{-80pt}
\caption{Simulation results showing anamorphism and its effect on the f\#
of a system consisting of a WP illuminated in a collimated beam, followed by a
lens with a focal length of 100~mm. The WP is tilted from 0--30$^{\circ}$ in 
steps of 3$^{\circ}$ and the ratio of beam diameters along $x$ and $y$-directions 
is measured to obtain the eccentricity of the beam for both E- and O-rays 
(left panel). The change in f\# due to the anamorphic beam is shown on the 
right. In order to minimise the change in f-numbers of the O- and E-rays, 
the WP must be tilted by around 24$^{\circ}$.}
\label{fig02b}
\end{figure*}

The lenses are indexed L1, L2, and L3, with the suffix `V' and `G' reserved for the 
first lens L1, when used in the VTT and GREGOR configurations, respectively.  All 
the lenses are off-the-shelf components from Edmund Optics with a diameter of 25~mm. 
The lenses L1V and L1G are achromatic doublets with a NIR II and MgF$_2$ coating, 
respectively. The lenses L2 and L3 are plano-convex, with a MgF$_2$ coating. All lenses 
have a scratch-dig of 40--20, as specified by the vendor. The average reflectivity of the 
lenses at 850~nm is less than 1.75\%. The lenses L2 and L3 are separated 
by a distance of 0.3~mm and de-centred from the optic axis by 2~mm. The WP is at a 
distance of 40 mm from the lens L1 and the f\# of the input beam to WP is more than 
350 (i.e., the beam is nearly collimated). Thus, the combined unit of 
WP--L2--L3--Detector is common to both the VTT and GREGOR configurations that can be 
mounted together, while the lens L1 can be a separate unit that can be interchanged 
between the two configurations.

The WP from Thorlabs uses scatter-free $\alpha-$BBO crystal as the substrate, with an 
extinction ratio greater than 100000:1, and a scratch-dig of 20--10. We selected this 
prism from different off-the-shelf components for the required beam separation in the 
image plane. The beam separation is 20$^\circ$ at 250~nm and 15$^\circ$ at 854~nm.
The wedge angle of the WP is rotated by 24.5$^\circ$ with respect to the 
$x$-axis and it produces a separation of 0.85~mm and 0.6~mm between the two orthogonal 
beams for the VTT and GREGOR configurations, respectively. The rotation 
of the WP affects the f\# of the O- and E-rays, thereby producing different magnifications
in the two beams. This effect can be attributed to anamorphism (Fig.~\ref{fig02b}), 
wherein a WP in normal incidence causes the beam to be compressed along one direction 
\citep{1986ApOpt..25..369S}. The effect of the rotation of the WP, as shown in 
Fig.~\ref{fig02} is to mitigate the above effect. A preliminary analysis of the design 
for polarization measurements using ZEMAX, shows that there will not be any cross-talk 
between the $s$ and $p$ polarization states, irrespective of the rotation of the WP. However, 
this rotation reduces the transmittance nominally. This analysis excludes the effect 
of coating of the WP surfaces.

\begin{figure*}[!ht]
\centerline{
\frame{\includegraphics[width=0.5\textwidth]{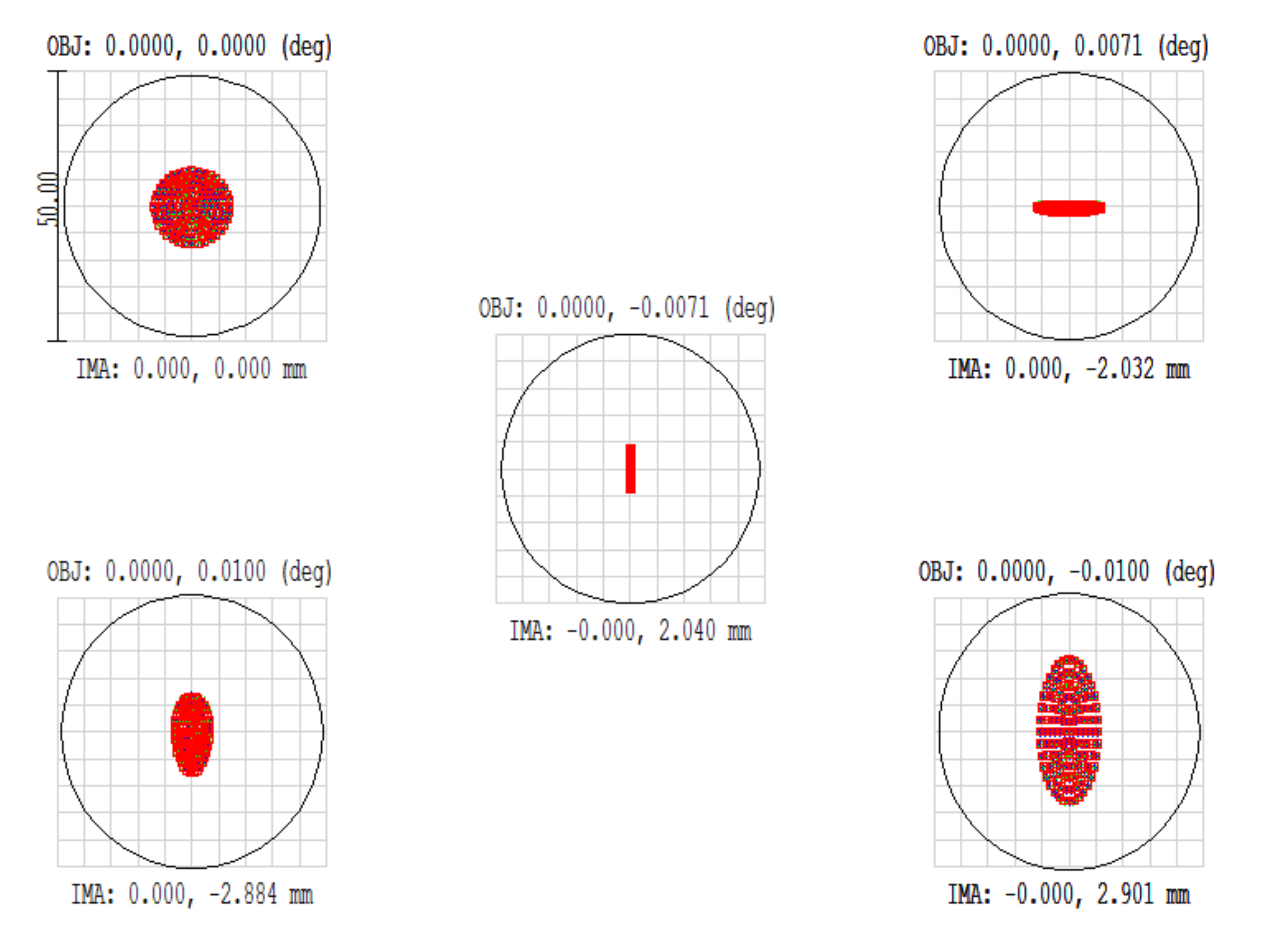}}
\frame{\includegraphics[width=0.5\textwidth]{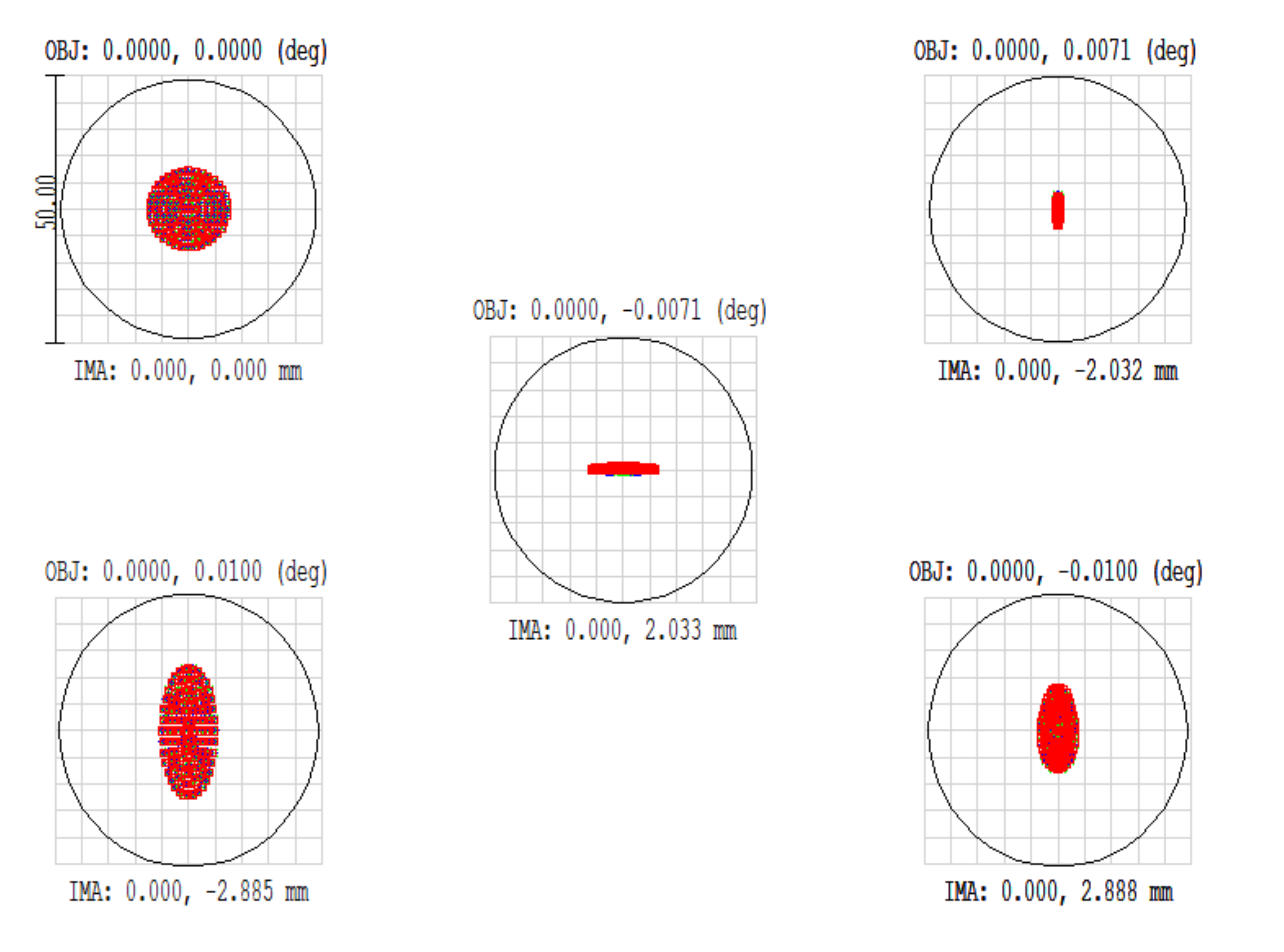}}
}
\caption{ZEMAX spot diagrams for the VTT configuration at 854~nm. The left and right panels 
correspond to the O- and E-rays, respectively.}
\label{fig03}
\vspace{10pt}
\centerline{
\frame{\includegraphics[width=0.5\textwidth]{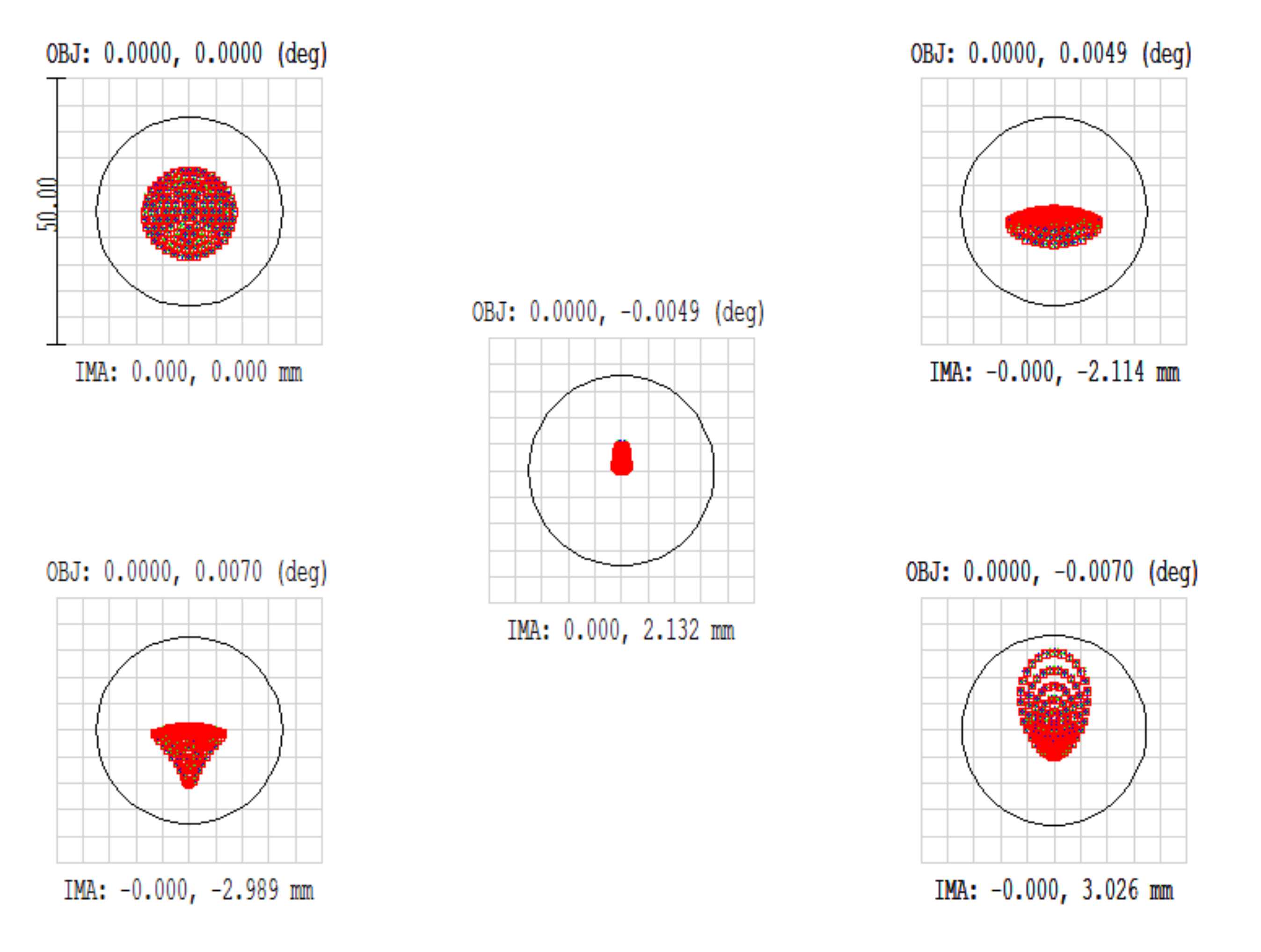}}
\frame{\includegraphics[width=0.5\textwidth]{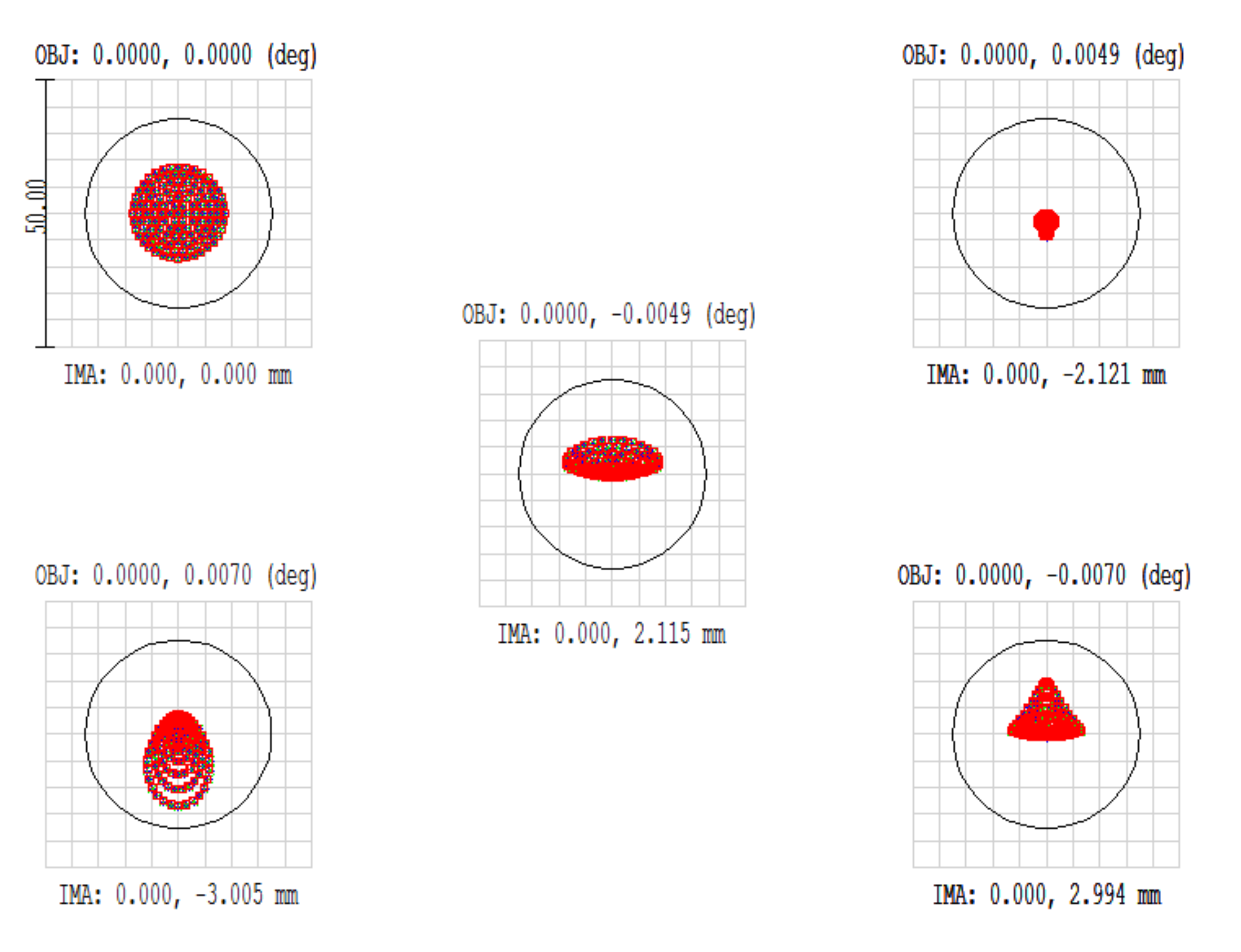}}
}
\caption{Same as Fig.~\ref{fig03} but for the GREGOR configuration.}
\label{fig04}
\end{figure*}

\begin{figure*}[!ht]
\centerline{
\includegraphics[angle=90,width=0.95\textwidth]{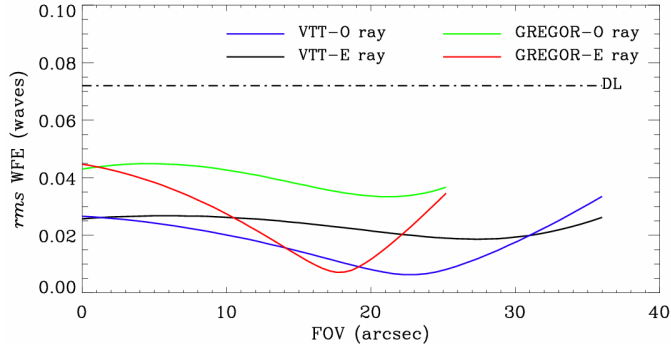}
}
\vspace{-50pt}
\caption{Variation of the $rms$ wavefront error with the FOV for the VTT and GREGOR 
configurations.}
\label{fig05}
\end{figure*}

\begin{figure}[!ht]
\vspace{10pt}
\centerline{
\includegraphics[angle=0,width=0.7\textwidth]{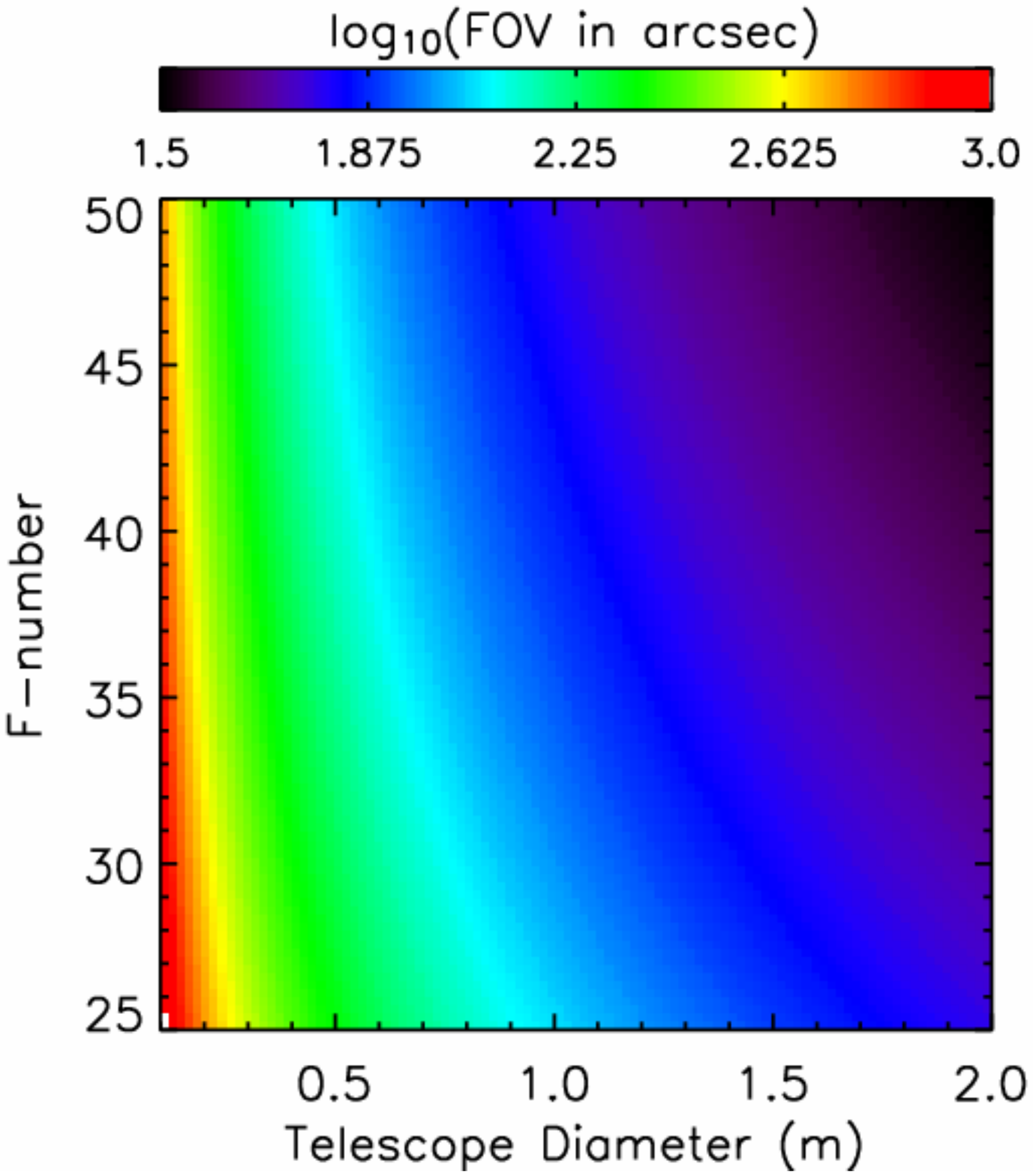}
}
\vspace{-35pt}
\caption{Permissible FOV for different telescope diameters and f-numbers that can be realised
with ACROPOL.}
\label{fig06}
\end{figure}

%%Table~\ref{tab01} summarises the 
%%various dimensions and the f\# conversion using ACROPOL at the VTT and GREGOR. 
%%Thus, one can obtain diffraction limited performance for f-numbers varying from 
%%33 to 23, and 40 to 17 for the VTT and GREGOR configurations, respectively, while 
%%maintaining a parfocal system. 
Figures~\ref{fig03} and \ref{fig04} show the spot 
diagrams for different points in the FOV for the O- and E-rays for both 
configurations. As is evident, all the aberrations are well 
within the Airy disc and the design can be considered as diffraction-limited. 
Figure~\ref{fig05} shows that the $rms$ wavefront error is well within  the diffraction 
limit, corresponding to a Strehl ratio of 0.82, for the FOV considered in the two 
configurations. The present optical setup would be useful for a range of input 
f-numbers varying from 25--40 and 30--50 for the VTT and GREGOR, respectively, with 
a change in the position of lens L1. The adaptability is also illustrated 
in Fig.~\ref{fig06} which shows the permissible FOV for different f-numbers 
and telescope diameters that satisfy the following two conditions, (i) the f\# of the 
telescope-spectrograph system (i.e. at SFP) should be larger than 25, and (ii) the image 
size at SFP should be less than 14 mm. Using the above conditions, we arrive at 
$\Theta\times F<48000$, where $\Theta$ is the FOV in arcmin and $F$ is focal length at SFP in mm.
Our proposed design is applicable for any FOV and f\# that satisfies the above condition, and 
will be nearly diffraction limited with a minor adjustment in the position of the first lens.
%%%although this will introduce a change in the 
%%%plate scale. 

We also performed a tolerance analysis for our design. We set the following criteria for
determining the tolerances, namely, the minimum beam separation should be 0.5~mm, the image 
size should be will within the detector size, the change in f\# between O and E-rays should be 
minimal, and the performance of the lens system should be diffraction limited. With these 
conditions, we obtain the following tolerances. The permissible 
changes in tilt and de-centre of the individual components are 30~\arcmin~and $\pm$1~mm, 
respectively. The distance between the components, L1--WP and WP--L2, can vary around 10\% of 
the actual value. The distance between L2 and L3 can be $\pm$0.2~mm. All these changes affect
the f\# of the system, but equally for both the O and E-rays. 
The tolerance for the WP rotation is $\pm$3$^\circ$, which causes a change in the f\# for 
both beams differentially. However, the consequent change in plate scales in less than 
0\farcs005. The above tolerances imply that the setup can be easily aligned without 
compromising the performance of the system.

It is to be noted that, by changing L1 and its position, the WP-L2-L3-Detector unit
can be utilised for another telescope with a different diameter. For 
example, using either L1V or L1G, and changing its position, the same setup 
can be used for a 1~m telescope with an f\# varying from 20--50, and a FOV of 
50\arcsec~to 80\arcsec.

\subsection{FLC Modulators}
The choice of FLC modulators for ACROPOL is straightforward, as the instrument is intended 
for a single wavelength operation, similar to GRIS. One of the FLCs serves as a half-wave 
retarder while the other as a quarter-wave retarder. The switching time of around 0.1~ms is 
achieved by changing the orientation of the fast axis of the FLCs using a low DC voltage. 
With two orientations of each FLC, four sets of measurements yield the four Stokes parameters 
\citep{Almeida1994,1999ASPC..184....3C,2007insp.book.....D}. The difference in orientation 
angles of the optical axis of the two states for both crystals is 45$^\circ$. For a set of 
four measurements, the half-wave and quarter wave FLCs have duty cycles of 25\% and 50\%, 
respectively. Standard, mounted FLCs from Meadowlark Optics Inc. have a clear aperture of 
17.8~mm and a thickness of 19.05~mm. Taking into account the design parameters described 
in Table~\ref{tab01} and the size of the FLCs, there will be no vignetting if the crystals 
are placed just before the instrument focal plane. Since the orientation angle of the optic 
axis is temperature dependent ($-0.3^\circ/^\circ$C), the FLCs will have to be enclosed 
in a temperature-controlled oven.

\subsection{Detector}
\begin{figure}[!h]
\centerline{
\includegraphics[angle=90,width=1.2\textwidth]{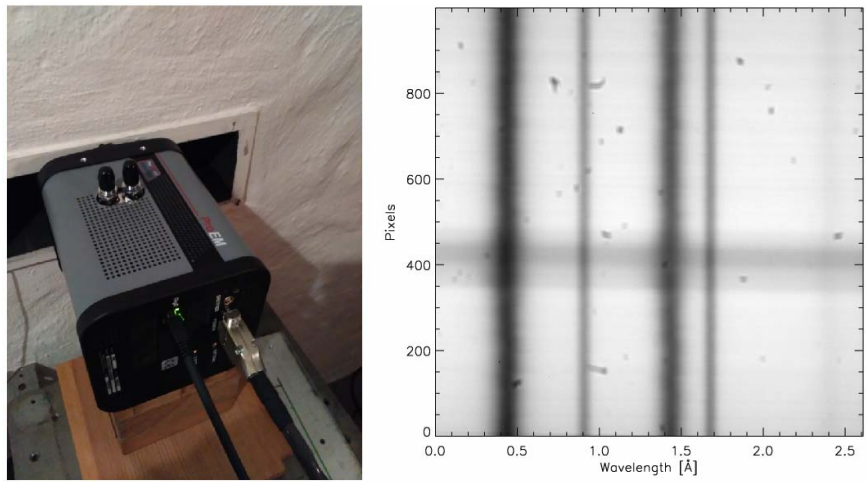}
}
\vspace{-85pt}
\caption{Left : ProEM CCD camera from Princeton Instruments at the focal plane 
of the Einstein Tower. Right: Unprocessed solar spectrum at 630~nm with an 
exposure time of 20~ms with the slit passing over a sunspot in NOAA AR 12548 
on 2016 May 27. Due to poor observing conditions, calibration data could not 
be taken that day.}
\label{fig07}
\end{figure}

A ProEM-HS:1024BX3 EMCCD camera from Princeton Instruments serves as the detector. 
The CCD has 1024$\times$1024~pixels of 13$\times$13~$\mu$m size. The sensor has a 
quantum efficiency of 95\% and 65\% at 650~nm and 850~nm, respectively. The camera 
features a readout speed of 30~MHz with the electron multiplication (EM) gain mode to 
deliver 25 frames/s, as well as a slow scan normal CCD readout mode with very low 
read noise for precision photometry applications. The dynamic range of the CCD is 
3900 at a readout rate of 30~MHz with 16 bit digitization. The sensor is coated 
with a patented anti-reflection coating for standard fringe suppression. At full 
resolution, the camera frame rate is given by $(t_{exp} + 0.04)^{-1}$ frames/sec 
with a readout rate of 30~MHz and a vertical shift rate of 700~ns, where $t_{exp}$ 
is the exposure time in seconds. The desired signal-to-noise ratio can be achieved 
by modifying the exposure time, the analog-to-digital rate, the vertical shift rate, 
and the EM gain. Other features of the camera include a Gigabit Ethernet interface 
for fast data transfer, proprietary data acquisition software, as well as software 
development kits for custom programming that are compatible with Linux and latest 
versions of Windows. The functionality of the camera was tested at the spectrograph 
of the Einstein Tower, Telegrafenberg, Potsdam and a sample spectrum at 630~nm is shown in 
Fig.~\ref{fig07}.

\section{Conclusions}
\label{conclu}
We present the design of a polarimeter to investigate the chromospheric 
magnetic field. The stand-alone instrument consists of three lenses, two ferro-electric 
liquid crystals, a Wollaston prism, and a CCD camera, allowing standard two-beam polarimetry. 
The consideration in the optical design was to enable the instrument to adapt to various 
f-numbers and telescope diameters and provide the necessary plate scale at the detector. To 
that extent, we describe the design for the 70~cm Vacuum Tower Telescope and the 1.5~m GREGOR 
solar telescope. In both cases, we obtain a diffraction-limited design, using off-the-shelf 
components, with an additional constraint that one of the lenses has to be replaced for 
the two configurations. However, the performance is valid over a range of 
f-numbers, so either configuration described here could be utilised for any existing
solar telescope. We plan to experimentally determine the effect of the rotation
of the Wollaston prism on the Muller matrices of the two orthogonal beams. However, since this
effect is expected is to be systematic for a fixed rotation angle, it can be corrected by a
suitable calibration scheme combining a quarter-wave plate and a linear polariser placed in front
of the polarimeter. The budget of the instrument is very modest, the bulk of the expense 
being the CCD camera. Its compactness and adaptability make it a valuable addition to 
any solar facility.

\texttt{Acknowledgements }
REL would like to thank Dr. J\"urgen Rendtel for his help in setting up the spectrograph 
at the Einstein Tower. A special note of thanks to Dr. Silke Schulze, from Roper Scientific 
Germany, for explaining the functionalities of the ProEM CCD camera and testing it at the 
spectrograph. We wish to thank Dr. Shibu Mathew for the discussion on the various 
materials used in Wollaston prisms. We would like to thank the technical
team at Thorlabs for providing us the ZEMAX files for the Wollaston prism.
We are grateful for the financial assistance from SOLARNET--the European 
Commission's FP7 Capacities Programme under Grant Agreement number 312495.
This work was also supported by grants AYA2014-60476-P and SP2014-56169-C6-2-R
at the Instituto de Astrof\'isica de Canarias, Tenerife, Spain. We thank the 
referees for their insightful comments.
%%``Solar Magnetometry in the Era of Large Solar Telescopes (AYA2014-60476-P)'' 
%%led by Drs. Héctor Socas-Navarro and Andrés Asensio Ramos, and project 
%%``Manufacturing and integration of the QM, FM, and FS models of 
%%SO/PHI (Polarimetric and Helioseismic Imager for Solar Orbiter)'' 
%%ESP2014-56169-C6-2-R, led by Dr. Basilio Ruiz Cobo.

%%\bibliographystyle{elsarticle-num}

%%\bibliographystyle{model1-num-names}

%% Numbered without titles
%\bibliographystyle{model1a-num-names}

%% Harvard
%\bibliographystyle{model2-names.bst}\biboptions{authoryear}

%% APA style
\bibliographystyle{model5-names}\biboptions{authoryear}

\bibliography{reference}

\end{document}